# Condition-aware learning enables robust prediction of oligonucleotide melting behavior across diverse chemistries and assay conditions

**Authors:** Danielle L. Ferreira PhD*, Lifeng Lin PhD, Adam Aslam, Nicholas Chang, Rebekah G. Baig, Edgar Baculi, Zoey Cao, Melanie Senn PhD

***Corresponding author:** Danielle.lopesferreiraastuto@cepheid.com

**Affiliation:** Cepheid, Sunnyvale, CA, USA.

**Institution:** Cepheid, Data Science, 904 Caribbean Drive, Sunnyvale, CA 94089, USA.

**Abbreviations:** AI = Artificial intelligence, DL = Deep learning, NT = Nucleotide Transformer (genomic foundational models), PCR = Polymerase chain reaction, qPCR = Quantitative PCR (real-time PCR), Tm = Melting temperature (°C), DNA = Deoxyribonucleic acid, , LNA = Locked nucleic acid, +T LNA = Oligonucleotide containing LNA-modified thymine, +C LNA = Oligonucleotide containing LNA-modified cytosine, GC% = Guanine–cytosine content (percentage of G+C bases in a sequence), NN = Nearest-neighbor (thermodynamic) model, $K^+$ = Potassium ion concentration, $Mg^{2+}$ = Magnesium ion concentration, $Na^+$ = Sodium ion concentration, MAE = Mean absolute error (°C), MSE = Mean squared error (°C$^2$), RMSE = Root mean squared error (°C).

## Abstract

Oligonucleotide melting temperature is a fundamental determinant of nucleic acid hybridization and underpins the design of molecular diagnostics, polymerase chain reaction assays, and many other biotechnology applications. However, accurately predicting melting behavior remains difficult because it depends not only on sequence composition, but also on experimental conditions and chemical modifications commonly used in modern assay design. Existing thermodynamic models rely on fixed parameterizations that are often difficult to extend across diverse reaction environments and nucleotide chemistries.

Here we show that a condition-aware nucleotide language model can accurately predict oligonucleotide melting behavior across diverse experimental conditions and both unmodified and chemically modified oligonucleotides. By combining contextual sequence representations with explicit information describing the reaction environment, the framework achieves sub-degree prediction accuracy and reduces prediction error for locked nucleic acid–modified oligonucleotides by up to 25% relative to nearest-neighbor thermodynamic approaches. The model also more accurately captures the thermal effects introduced by nucleotide modification and maintains strong performance on independent benchmark datasets spanning experimental conditions substantially different from those represented during training.

Our results demonstrate that learned sequence representations can complement classical thermodynamic models by capturing context-dependent effects that are difficult to encode using fixed parameter tables alone. More broadly, this work provides a scalable framework for predicting oligonucleotide melting behavior across diverse chemistries and assay conditions, supporting more reliable molecular assay design.

# Introduction

Accurate prediction of nucleic acid duplex stability remains a central challenge in molecular biology and biotechnology, particularly as molecular assays increasingly employ diverse buffer compositions and chemically modified nucleotides. Melting temperature ($T_m$) is a widely used measure of hybridization stability and underpins the design and performance of numerous applications, including PCR, digital PCR, targeted sequencing, and molecular diagnostics[1–4]. As assay platforms demand higher precision and multiplexing capability, reliable $T_m$ prediction across heterogeneous experimental conditions has become increasingly critical[3,5].

The prevailing computational framework for $T_m$ prediction is the nearest-neighbor (NN) thermodynamic model, which estimates duplex stability from the additive contributions of adjacent nucleotide pairs[6,7]. Although highly successful for canonical DNA duplexes, NN models depend on experimentally derived parameter tables and assume that local interactions contribute independently to overall stability. Extensions to account for salt, oligonucleotide concentration, and other reaction variables are applied as empirical corrections, introducing compounding sources of error outside the original calibration regime[7]. More fundamentally, the NN formulation scales poorly as sequence, chemical, and environmental complexity increase: each new nucleotide modification or interaction context requires explicit enumeration of additional parameters, rapidly expanding model dimensionality while remaining constrained by assumptions of locality and linearity. As reflected in curated resources such as the Nucleic Acid Thermodynamic Database [17], comprehensive parameter coverage across diverse chemistries and assay conditions remains incomplete.

These limitations are particularly relevant for locked nucleic acid (LNA) oligonucleotides. LNAs are widely used in molecular diagnostics and nucleic acid detection because they generally increase duplex affinity and specificity[8]. However, the thermodynamic impact of LNA incorporation depends on modification type, sequence context, position within the oligonucleotide, and experimental conditions. Existing NN-based approaches rely on relatively small sets of experimentally derived correction parameters and therefore struggle to capture the full diversity of modification-dependent effects encountered in practical assay design[7]. As a result, accurate prediction of LNA-containing oligonucleotides remains a persistent challenge despite their widespread use.

Recent advances in machine learning offer an alternative strategy for modeling nucleic acid thermodynamics. Rather than relying on predefined thermodynamic parameters, data-driven approaches can learn relationships directly from experimental observations, potentially capturing nonlinear sequence dependencies and context-specific interactions that are difficult to encode explicitly[9,10]. At the same time, large-scale nucleotide language models trained on genomic sequences have demonstrated an ability to learn biologically meaningful sequence representations that transfer effectively to downstream prediction tasks[11]. Unlike hand-engineered sequence descriptors, nucleotide language models learn contextual representations that encode information from the surrounding sequence environment. These representations can capture higher-order sequence relationships and long-range dependencies that are difficult to express through nearest-neighbor parameterizations. When combined with experimental training data, they can also learn context-dependent effects associated with nucleotide modifications and assay conditions. Such properties make them a promising foundation for predicting oligonucleotide melting behavior across diverse chemistries and experimental environments. However, existing learning-based approaches have largely focused on unmodified nucleic acids and have typically been developed under relatively narrow experimental conditions, limiting their applicability to chemically modified probes and diverse assay environments[9,10].

Here, we present a condition-aware, data-driven framework for predicting oligonucleotide melting behavior across diverse sequence compositions, buffer conditions, and nucleotide chemistries. We combine large-scale in-silico thermodynamic simulations with targeted experimental measurements spanning a wide range of ionic environments and oligonucleotide concentrations representative of practical assay conditions. Using a dataset comprising both unmodified and LNA-modified oligonucleotides measured across a wide range of potassium, magnesium, and oligonucleotide concentrations, we train a model that directly predicts $T_m$ while remaining adaptable to experimentally relevant assay conditions.

We show that the proposed framework consistently outperforms nearest-neighbor thermodynamic models, conventional machine-learning approaches, and widely used design tools across both unmodified and LNA-modified oligonucleotides. Beyond improving absolute $T_m$ prediction, the model more accurately captures modification-induced thermal effects ($\Delta T_m$) between matched modified and unmodified probes, indicating improved modeling of context-dependent LNA behavior. Furthermore, the framework maintains strong performance on independent public benchmarks spanning ionic and concentration regimes well beyond those represented during training, demonstrating robust out-of-distribution generalization. Together, these results highlight the potential of condition-aware nucleotide language models as a scalable alternative to fixed thermodynamic parameterizations for oligonucleotide design and analysis.

# Results

## AI framework for melting temperature prediction

We developed a condition-aware deep learning framework to predict oligonucleotide melting temperature ($T_m$) under specified reaction conditions. In this context, *condition-aware* refers to the model's ability to explicitly incorporate assay variables into the prediction process, rather than relying on fixed post-hoc corrections.

As illustrated in Figure 1, the proposed framework integrates a pretrained nucleotide language model with condition-aware adaptation to predict oligonucleotide $T_m$. Assay conditions, including potassium ($K^+$), magnesium ($Mg^{2+}$), and oligonucleotide concentration, are encoded alongside the sequences of the hybridizing oligonucleotides and processed by the Nucleotide Transformer v2-250M encoder. The pretrained encoder is kept frozen, while task-specific adaptation is achieved through low-rank adaptation (LoRA) modules. The adapted representations are subsequently analyzed by two complementary prediction branches: a global regression head that captures sequence-wide and assay-condition-dependent determinants of melting temperature, and local residual heads that model position-specific neighborhood effects, including effects associated with modified bases when present. The local contributions are aggregated and incorporated as residual corrections to the global prediction, enabling the model to integrate broad sequence context, assay conditions, and localized thermodynamic effects. This design combines the rich representations learned during large-scale pretraining with architectural components motivated by nucleic acid thermodynamics.

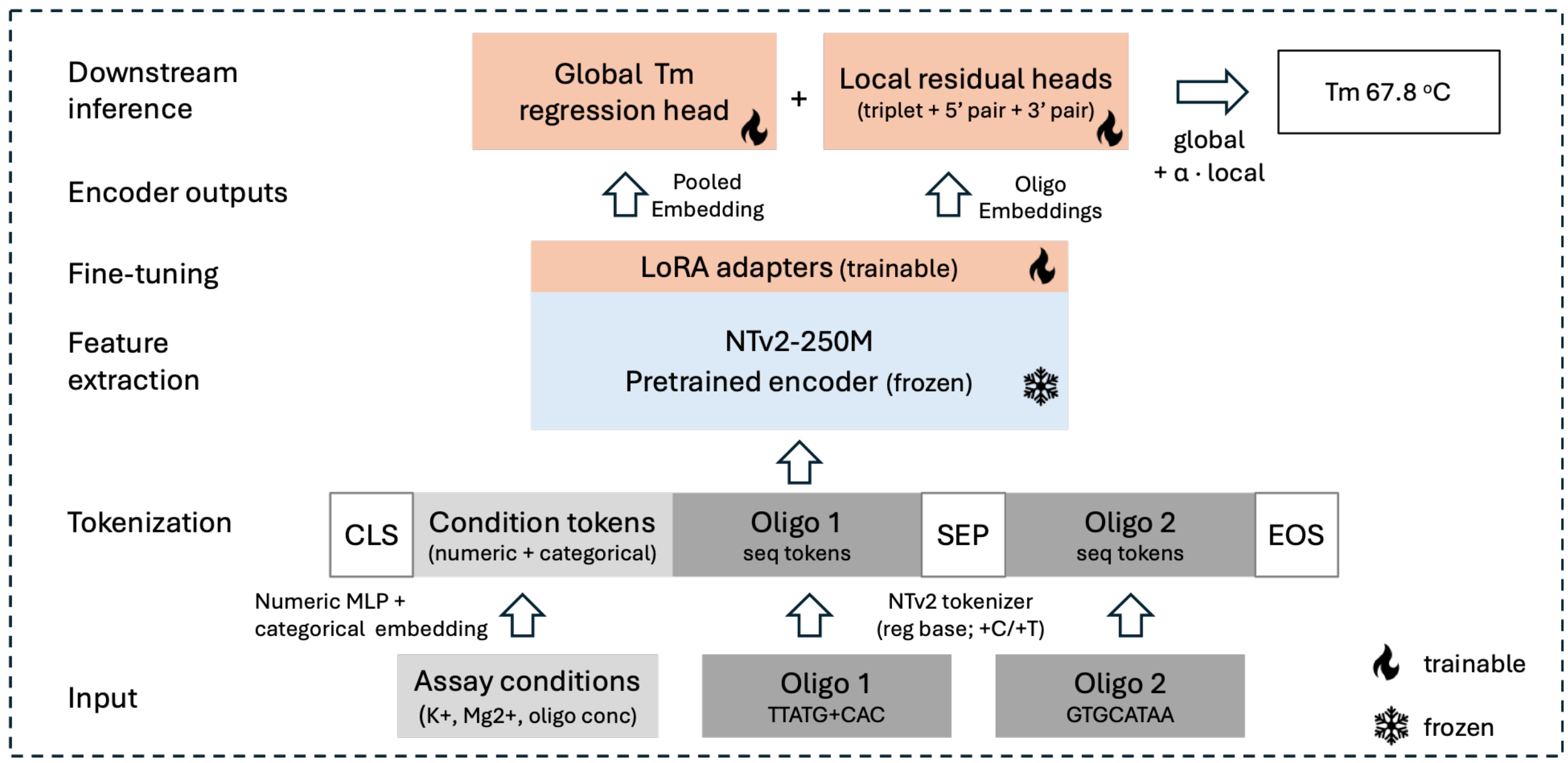

**Figure 1. AI framework for Melting Temperature predictions.** Condition tokens representing assay conditions ($K^+$, $Mg^{2+}$, and oligonucleotide concentration) are combined with the sequences of the two hybridizing oligonucleotides and processed by a pretrained Nucleotide Transformer adapted using LoRA modules. A global regression head and local residual heads jointly generate the final Tm prediction, with local contributions aggregated as residual corrections to the global estimate.

The training corpus consisted of three related datasets derived from a common pool of unmodified oligonucleotide pairs: Regular bases, LNA+C, and LNA+T (Table 1). The base dataset included melt-curve measurements for approximately one thousand duplexes, each representing a unique target sequence paired with its perfect reverse-complement strand (serving as a binding target). From this base set, two additional LNA-modified datasets were generated by introducing +T or +C substitutions into one strand of each duplex, while leaving the complementary strand unaltered, mimicking the modified oligo binding to an unmodified natural target sequence. This design produced two distinct derivative collections, each reflecting the hybridization behavior of a modified probe binding to an unmodified target.

The number of modified positions varied across sequences, and the resulting +T and +C datasets differed in size and composition accordingly. All three datasets were evaluated across a matrix of sodium and magnesium concentrations and multiple oligonucleotide concentrations (Table 1), yielding a comprehensive set of melt-curve–derived Tm values. Model training incorporated an initial pretraining step on a large synthetic dataset, followed by fine-tuning on the experimental measurements (see Methods for details of synthetic data generation). Together, these datasets enabled the model to learn hybridization behavior for both unmodified and LNA-modified oligos under conditions representative of real assay environments.

**Table 1. Summary statistics of the experimental oligonucleotide melting temperature dataset.**

| Category | Statistic | Total (train \| val \| test splits) |
|---|---|---|
| **Dataset composition Measurements** | Total measurements | 5180 (3607 \| 516 \| 1057) |
| | Regular bases | 690 (631 \| 99 \| 196) |
| | LNA-C variants | 2304 (1491 \| 211 \| 441) |
| | LNA-T variants | 2304 (1426 \| 206 \| 420) |
| **Experimental Tm** | Range (Train + val \| test) | 52–83 \| 54-78°C |
| | Mean ± SD (train + val \| test) | 68.4 ± 4.2 \| 68.2 ± 4.4°C |
| **ΔTm (replicates)** | Median (IQR) | 0.10 (IQR 0.00–0.20 °C) |
| **Sequence length** | Range (Train + val \| test) | 20–30 \| 20-30 nt |
| **GC content** | Range (Train + val \| test) | 22–77 \| 24-74 % |
| **Oligo concentration** | Range (Train + val \| test) | 100–1150 \| 250-800 nM |
| **Monovalent salt ($K^+$) concentration** | Range (Train + val \| test) | 50-500 \| 50-500 mM |
| **Divalente ($Mg^{2+}$) concentration** | Range (Train + val \| test) | 0-11 \| 0-11 mM |

Melting temperatures are reported as the mean of duplicate measurements, with ΔTm indicating replicate variability.

## Regression Performance and Benchmark

Performance was evaluated on a held-out test set comprising 1,057 oligonucleotide-target pairs, whereas model development and hyperparameter optimization were performed exclusively on a separate training pool of 4,123 pairs (Table 1). The proposed AI framework was compared against *Thermo*, a nearest-neighbor thermodynamic baseline with salt correction[6],[7] as well as conventional machine-learning approaches (linear regression and XGBoost). To ensure a fair comparison, *Thermo* and both machine-learning baselines were fitted on the same training split used for AI-model development and evaluated on identical test set. Additional comparisons were performed against widely used oligonucleotide-design tools where applicable. Quantitative performance metrics are summarized in Table 2, whereas predicted-versus-experimental regression analyses are shown in Figures 2 and 3.

**Performance on LNA-modified oligonucleotides.** Because LNA-modified oligonucleotides represent the primary focus of this study, we first evaluated predictive performance on the LNA+T and LNA+C datasets. Across both modification classes, the AI model achieved the strongest agreement with experimental measurements and the lowest prediction errors (Table 2). For LNA+T oligonucleotides, the AI model achieved an MAE of 0.79 °C and $R^2$ of 0.94, compared with 1.06 °C and 0.91 for *Thermo*. Similar improvements were observed for LNA+C oligonucleotides, where the AI model achieved an MAE of 0.80 °C and $R^2$ of 0.95, whereas *Thermo* yielded an MAE of 1.00 °C and $R^2$ of 0.92. Regression analyses further demonstrated tighter agreement with experimental measurements across the full temperature range (Figure 2). Overall, Pearson correlations exceeded 0.97 for both LNA datasets, indicating robust prediction accuracy for chemically modified oligonucleotides.

To investigate whether prediction accuracy depended on modification placement, we stratified $T_m$ prediction errors according to the position of the LNA within the oligonucleotide (Supplementary Table S2). The thermodynamic baseline *Thermo* exhibited larger errors and increased positive bias for terminal and near-terminal modifications, whereas the AI model maintained comparatively stable accuracy across position classes. The largest performance gains were observed for terminal modifications, for which the AI model reduced prediction error by up to 41% relative to the thermodynamic baseline. These findings suggest that the proposed framework more effectively captures position-dependent effects of LNA incorporation, particularly near oligonucleotide termini.

**Performance on unmodified oligonucleotides.** The AI model also achieved the best performance on unmodified oligonucleotides. For regular bases, the AI model achieved an MAE of 0.76 °C, compared with 0.86 °C for *Thermo*, while maintaining minimal bias and strong agreement with experimental measurements (Table 2; Figure 3). Although the performance gap relative to *Thermo* was smaller for unmodified sequences than for LNA-containing probes, the AI model consistently produced lower error and narrower limits of agreement.

**Comparison with conventional machine-learning approaches.** To determine whether these gains arose primarily from the deep sequence representation rather than supervised learning alone, we compared the proposed framework against linear regression and XGBoost models trained on the same data. Both conventional machine-learning approaches showed substantially lower predictive performance. For LNA-containing oligonucleotides, MAEs increased from 0.79-0.80 °C for the AI model to 1.39-1.60 °C for linear regression and XGBoost, respectively, accompanied by marked reductions in both Pearson correlation and $R^2$ (Table 2).

**Comparison with thermodynamic tools.** We additionally benchmarked the proposed framework against commonly used oligonucleotide-design software. Primer3, which does not support LNA chemistry and was not retrained on the study dataset, exhibited systematic overestimation of melting temperatures and larger prediction errors than the AI model (Figure 3; Table 2). Visual OMP likewise showed substantial positive bias and markedly higher absolute error. Although these external tools were evaluated using their default parameterizations rather than being retrained on the study data, they provide a practical reference against software commonly used in oligonucleotide-design workflows.

Collectively, these results demonstrate that the proposed condition-aware AI framework consistently outperforms thermodynamic, conventional machine-learning, and software-based approaches for $T_m$ prediction. The largest performance gains were observed for LNA-modified oligonucleotides, highlighting the value of contextual nucleotide representations for modeling the sequence-dependent effects of chemical modifications. We next examined the robustness of these predictions across temperature ranges, sequence compositions, and experimental conditions.

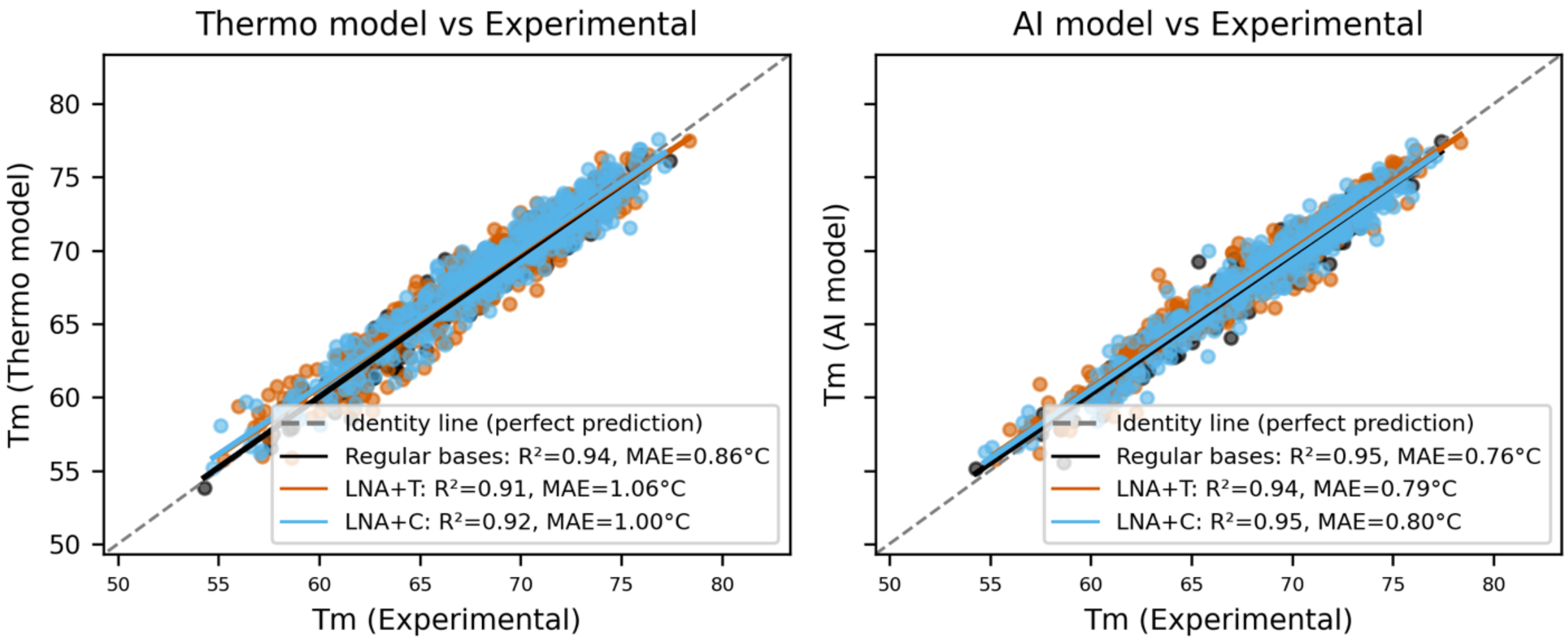

**Figure 2. AI versus thermodynamic modeling.** Predicted versus experimental melting temperatures ($T_m$) for the thermodynamic baseline (*Thermo*) and the proposed AI model on the held-out test set. Colors indicate oligonucleotide classes (regular, LNA+T, and LNA+C). Dashed lines denote perfect agreement (y=x). The AI model exhibits tighter agreement with experimental measurements across all classes, particularly for LNA-modified oligonucleotides.

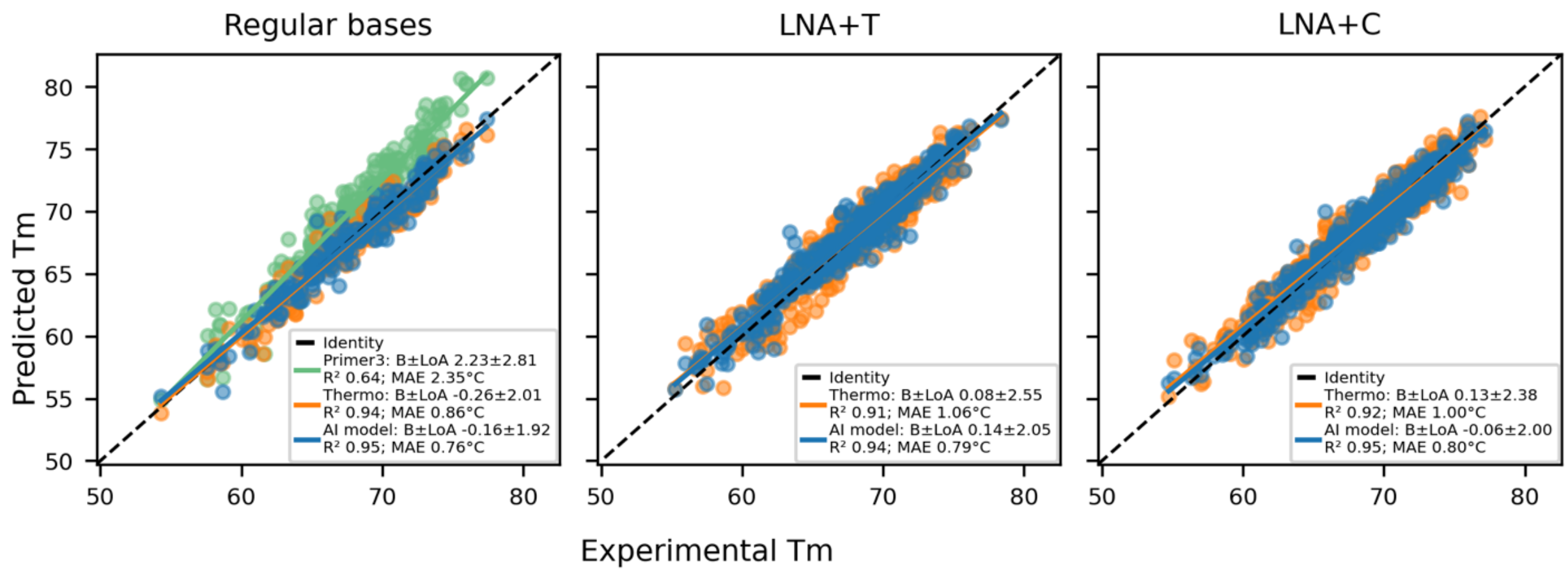

**Figure 3. Benchmarking against existing prediction tools.** Predicted versus experimental melting temperatures (Tm) for *Thermo*, Primer3 (regular oligonucleotides only), and the proposed AI model. Dashed lines indicate perfect agreement (y=x). Solid lines represent least-squares regression fits. Comprehensive performance metrics are reported in Table 2.

**Table 2. Performance, correlation, and agreement metrics for $T_m$ prediction on the held-out test set.**

| Dataset | Method | Category | Bias ± LoA ↓ | MAE [CI] (°C) ↓ | RMSE [CI] (°C) ↓ | MSE (°C²) ↓ | R² ↑ | Pearson r ↑ | p-value ↑ |
|---|---|---|---|---|---|---|---|---|---|
| LNA+T | **AI model** | Proposed | **0.14 ± 2.05** | **0.79 [0.73, 0.86]** | **1.06 [0.96, 1.16]** | **1.11** | **0.94** | **0.97** | 0.00170 |
| | **Thermo** | Dataset-trained baseline | 0.08 ± 2.55 | 1.06 [0.99, 1.13] | 1.30 [1.22, 1.38] | 1.69 | 0.91 | 0.96 | 0.15590 |
| | **LR** | | 0.37 ± 3.41 | 1.39 [1.29, 1.50] | 1.78 [1.65, 1.92] | 3.16 | 0.84 | 0.92 | 0.00001 |
| | **XGBoost** | | 0.54 ± 3.68 | 1.54 [1.42, 1.66] | 1.95 [1.81, 2.10] | 3.82 | 0.8 | 0.91 | 0.00000 |
| | **OMP†** | External software | 11.24 ± 6.30 | 11.24 [10.9, 11.6] | 11.69 [11.3, 12.0] | 136.62 | -4.96 | 0.92 | 8.1E-49 |
| LNA+C | **AI model** | Proposed | **-0.06 ± 2.00** | **0.80 [0.74, 0.86]** | **1.02 [0.94, 1.10]** | **1.04** | **0.95** | **0.97** | 0.15850 |
| | **Thermo** | Dataset-trained baseline | 0.13 ± 2.38 | 1.00 [0.93, 1.06] | 1.22 [1.15, 1.29] | 1.49 | 0.92 | 0.96 | 0.02290 |
| | **LR** | | 0.32 ± 3.82 | 1.54 [1.43, 1.65] | 1.97 [1.84, 2.11] | 3.9 | 0.8 | 0.9 | 0.00661 |
| | **XGBoost** | | 0.45 ± 3.95 | 1.60 [1.49, 1.71] | 2.06 [1.91, 2.21] | 4.26 | 0.78 | 0.89 | 0.00001 |
| | **OMP†** | External software | 9.60 ± 5.66 | 9.62 [9.30, 9.94] | 10.03 [9.71, 10.3] | 100.52 | -4.39 | 0.94 | 8.6E-49 |
| Regular bases | **AI model** | Proposed | **-0.16 ± 1.92** | **0.76 [0.67, 0.85]** | **0.99 [0.87, 1.12]** | **0.98** | **0.95** | **0.98** | 0.02920 |
| | **Thermo** | Dataset-trained baseline | -0.26 ± 2.01 | 0.86 [0.77, 0.95] | 1.06 [0.95, 1.16] | 1.12 | 0.94 | 0.97 | 0.00030 |
| | **LR** | | 0.20 ± 2.97 | 1.18 [1.05, 1.32] | 1.52 [1.37, 1.67] | 2.32 | 0.88 | 0.94 | 0.14290 |
| | **XGBoost** | | 0.41 ± 3.32 | 1.42 [1.28, 1.56] | 1.74 [1.58, 1.89] | 3.03 | 0.85 | 0.92 | 0.00215 |
| | **OMP†** | External software | 6.96 ± 4.73 | 6.99 [6.72, 7.26] | 7.36 [7.09, 7.64] | 54.24 | -1.34 | 0.96 | 9.1E-49 |
| | **Primer3†** | | 2.23 ± 2.81 | 2.35 [2.18, 2.52] | 2.65 [2.48, 2.81] | 7.03 | 0.64 | 0.97 | 4.5E-31 |

Metrics include MAE, RMSE, MSE, $R^2$, Pearson correlation, Bland-Altman bias and limits of agreement (LoA),

and Wilcoxon signed-rank test p-values. LR = Linear Regression. † External software benchmark; these methods were evaluated using their default parameterizations and were not retrained on the study dataset.

## Agreement analysis (Bland–Altman)

To assess prediction-experiment agreement beyond correlation and absolute error metrics, we performed Bland-Altman (BA) analysis across datasets, sequence characteristics, and assay conditions (Figures 4-5). BA analysis evaluates agreement by plotting the difference between predicted and experimental melting temperatures (predicted − experimental $T_m$) against their mean value, thereby quantifying both systematic bias and the spread of residuals through the mean bias and 95% limits of agreement (LoA).

Across all oligonucleotide chemistries, both the AI model and the thermodynamic baseline exhibited biases close to zero, indicating limited systematic over- or underestimation of $T_m$ values (Figure 4; Table 2). However, the AI model consistently achieved narrower limits of agreement, reflecting lower residual variability. For regular oligonucleotides, the AI model reduced the LoA from ±2.01 °C to ±1.92 °C. Larger improvements were observed for LNA-modified oligonucleotides, where the LoA decreased from ±2.55 °C to ±2.05 °C for LNA+T sequences and from ±2.38 °C to ±2.00 °C for LNA+C sequences.

To evaluate the robustness of these observations, we further stratified Bland-Altman analyses by GC content, sequence length, oligonucleotide concentration, potassium concentration, and magnesium concentration (Supplementary Figs. S1-S2). Across all strata, the AI model maintained narrower limits of agreement than the thermodynamic baseline, indicating that the reduction in residual variability is preserved across diverse sequence compositions and assay conditions

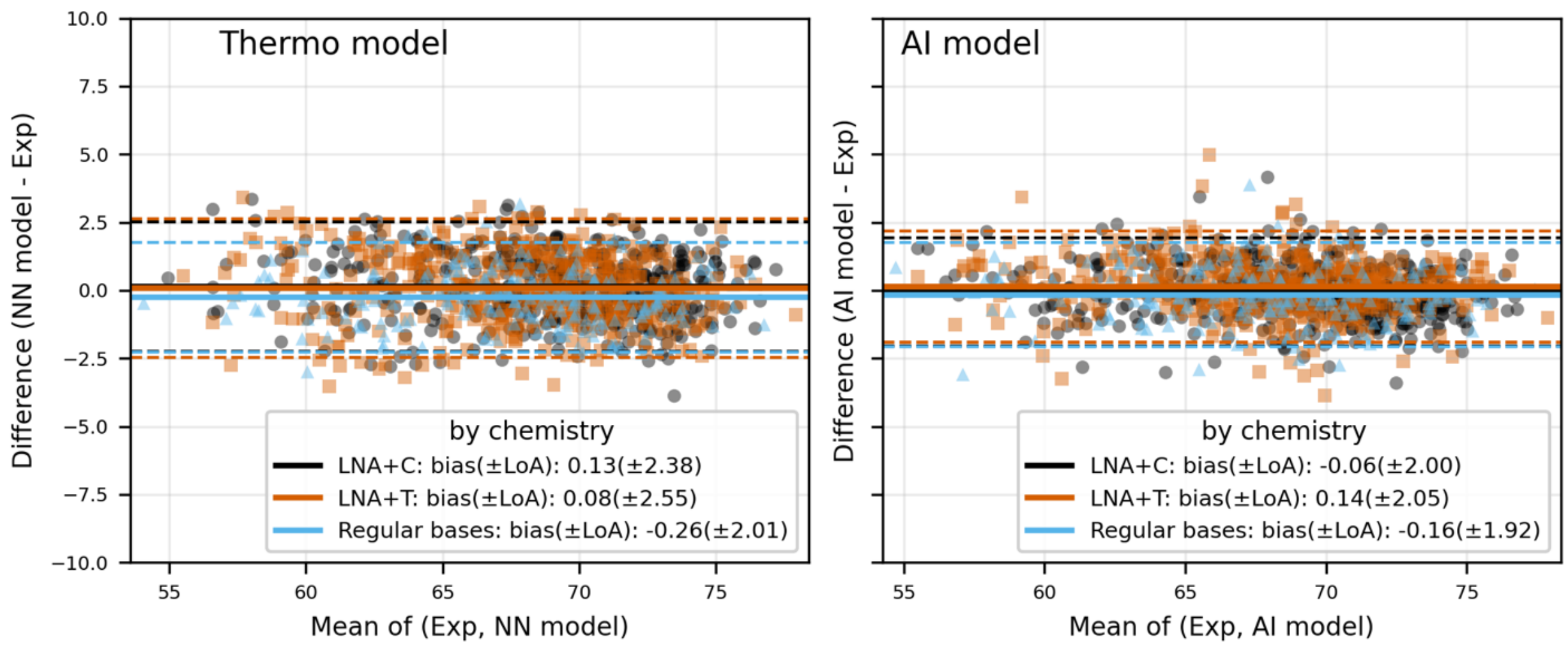

**Figure 4. Agreement analysis by oligonucleotide chemistry.** Bland-Altman plots comparing experimental $T_m$ with predictions from the thermodynamic baseline (Thermo) and the proposed AI model by dataset. Solid lines indicate mean bias and dashed lines indicate the 95% limits of agreement (LoA). Across all chemistry, the AI model exhibits narrower limits of agreement and reduced residual variability than the thermodynamic baseline.

## Prediction of modification-induced thermal effects

To evaluate whether the models could accurately capture the thermodynamic effect of LNA incorporation itself, we compared predicted and experimental ΔTm values derived from matched modified and unmodified oligonucleotide pairs. This analysis was performed on 367 paired measurements sharing the same target sequence and experimental conditions, with ΔTm defined as the difference between the melting temperatures of the modified and corresponding unmodified oligonucleotides.

The proposed AI model most accurately reproduced the observed $\Delta T_m$ values (Table 3). Compared with the thermodynamic baseline, the AI model reduced the prediction error by approximately 30% (MAE: 0.76 °C versus 1.11 °C) and more than doubled the explained variance ($R^2$: 0.63 versus 0.27). OMP exhibited substantially poorer agreement with experimental measurements.

Consistent with these findings, the AI model showed the smallest systematic bias and the narrowest limits of agreement.

Notably, the performance gap between methods was larger for ΔTm prediction than for absolute $T_m$ prediction. Whereas *Thermo* provided reasonable estimates of absolute melting temperature, its ability to predict the magnitude and direction of modification-induced thermal changes was substantially reduced. These findings indicate that the proposed framework more effectively captures the context-dependent effects of LNA incorporation, which are only partially represented by conventional nearest-neighbor thermodynamic models.

**Table 3. Prediction performance for modification-induced thermal effects ($\Delta T_m$) on 367 matched modified/unmodified oligonucleotide pairs.**

| Method | Bias ± LoA ↓ | MAE (°C) ↓ | RMSE (°C) ↓ | $R^2$ ↑ | Pearson r ↑ |
|---|---|---|---|---|---|
| AI model | **0.24 ± 1.79** | **0.76** | **0.95** | **0.63** | **0.79** |
| Thermo | 0.50 ± 2.52 | 1.11 | 1.38 | 0.27 | 0.52 |
| OMP† | 3.88 ± 3.86 | 3.88 | 4.25 | 0.07 | 0.27 |

## External validation and out-of-distribution generalization

While the preceding analyses evaluate performance on held-out data drawn from the same experimental collection, practical oligonucleotide design requires robust generalization to external datasets generated under different protocols, sequence compositions, and assay conditions. We therefore evaluated the proposed framework on two independent public benchmarks that were not used during model development.

**Evaluation on the *Panjkovich–Owczarzy* Benchmark.** To assess generalization under substantially different experimental conditions, we evaluated the model on the Panjkovich–Owczarzy Tm benchmark[12], a curated collection of 108 experimentally characterized DNA duplexes assembled from the studies of Owczarzy et al.[7,13] and Chiu et al.[14] This benchmark spans a broad range of ionic conditions and oligonucleotide concentrations and has been widely used for the development and evaluation of thermodynamic Tm prediction methods.

Previous evaluations on this benchmark reported MAEs ranging from approximately 12 °C for simple empirical formulas to 2.8 °C for consensus nearest-neighbor approaches (Figure 5). On this benchmark, the proposed AI model achieved an MAE of 2.04 °C, outperforming the fitted *Thermo* baseline (2.32 °C) and approaching the performance of Primer3 (1.36 °C) and the physics-based model proposed by Khandelwal et al. (1.33 °C) (Figure 5). Although the AI model did not achieve the lowest error on this benchmark, it remained competitive with methods that were developed and evaluated using datasets contributing to this benchmark.

Regression analysis further confirmed strong agreement between predicted and experimental melting temperatures across the full temperature range (Figure 6). The AI model achieved an $R^2$ of 0.95 despite being evaluated under conditions that differed substantially from those represented in the training data. In particular, the benchmark includes potassium concentrations ranging from 69 to 1,020 mM and oligonucleotide concentrations extending to 100,000 nM, whereas the training data spanned 100-500 mM $K^+$ and 100-1,150 nM oligonucleotide concentration. Despite this substantial distribution shift, the AI model maintained strong agreement with experimental measurements and performance comparable to established thermodynamic and physics-based approaches (Figure 6).

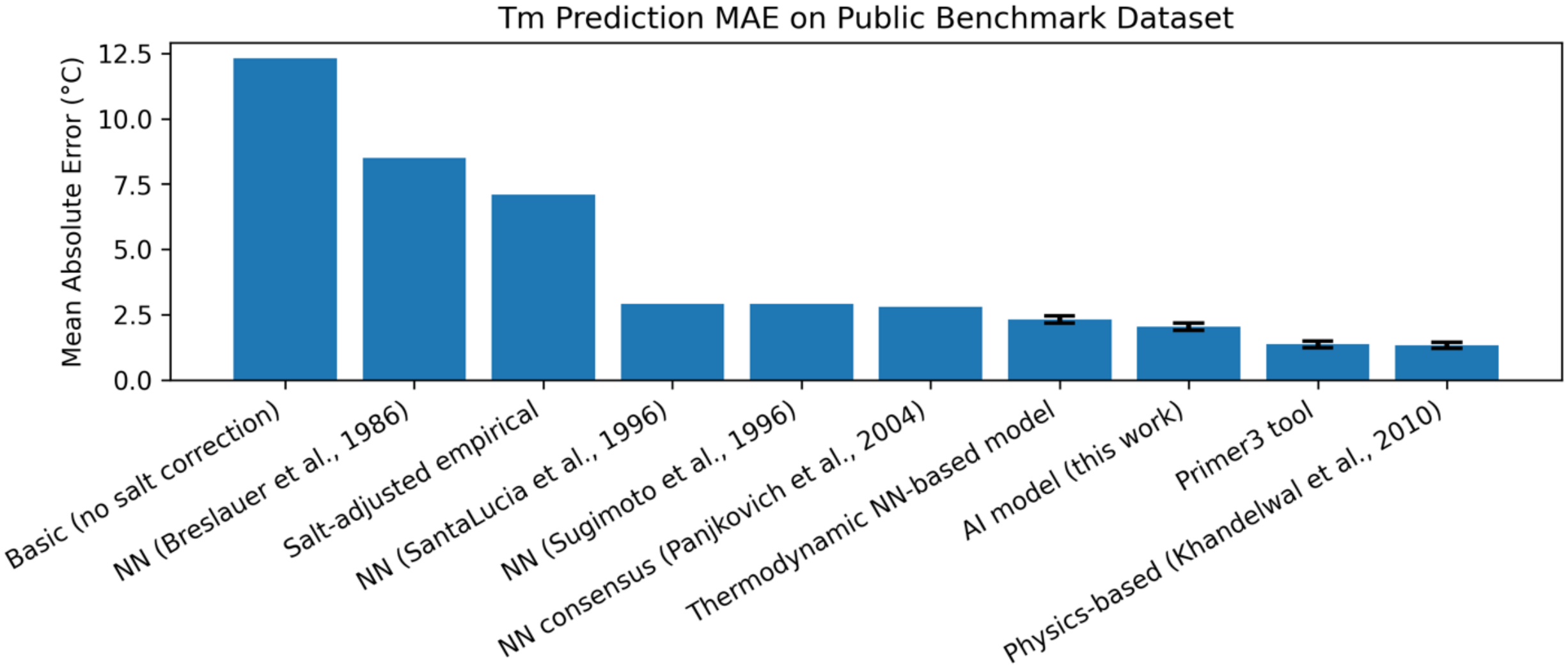

**Figure 5. Historical and contemporary performance on the Panjkovich–Owczarzy benchmark.** Mean absolute error (MAE) comparison for empirical, nearest-neighbor, physics-based, and AI-based Tm prediction methods. Earlier empirical and NN approaches exhibit substantially larger errors, whereas modern methods achieve MAEs below 2.5 °C.

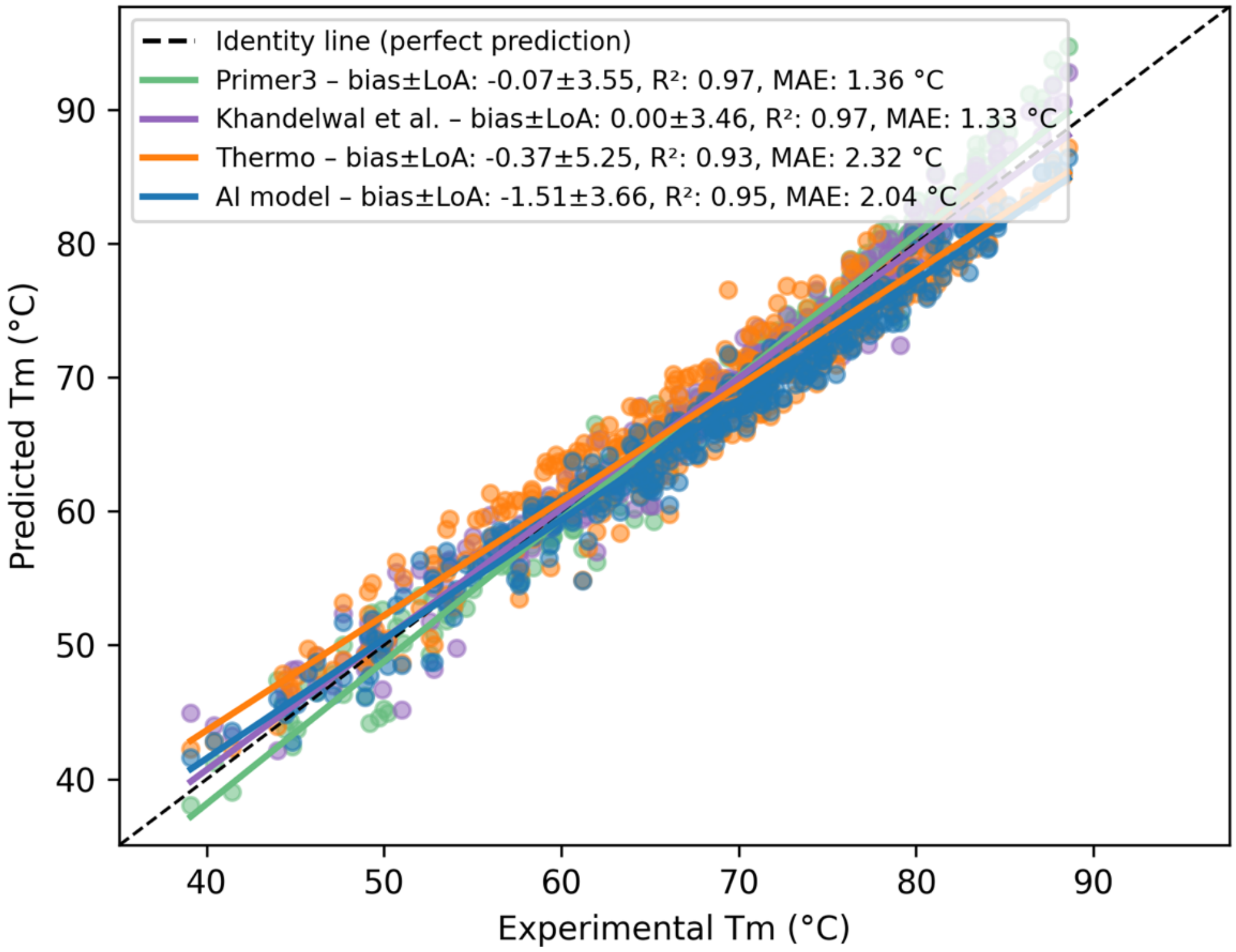

**Figure 6. Out-of-distribution performance on the Panjkovich–Owczarzy benchmark.** Predicted versus experimental melting temperatures for Primer3, the physics-based model of Khandelwal et al., the fitted thermodynamic baseline (*Thermo*), and the proposed AI model. All methods exhibit strong agreement with experimental measurements despite the broad range of ionic conditions represented in the benchmark.

**Evaluation on the Bakhtiarizadeh Benchmark.** We next evaluated the model on the Bakhtiarizadeh[15] benchmark, which comprises 79 experimentally validated primer pairs and was originally developed to compare 22 primer-design tools. Unlike our training and test datasets, this benchmark reports experimentally determined annealing temperatures (Ta) rather than melting temperatures (Tm), introducing an expected offset between predictions and measurements.

Because Ta is typically several degrees (commonly 3–5 °C) below Tm in PCR assay design[16], absolute bias should be interpreted in this context. We therefore evaluated agreement using Bland-Altman analysis (Figure 7). Among all evaluated methods, the proposed AI model exhibited the narrowest limits of agreement while maintaining a positive bias (+3.65 °C) consistent with the expected thermodynamic relationship between Tm and Ta. In contrast, many existing tools displayed biases clustered around zero or below zero relative to Ta, suggesting different calibration assumptions.

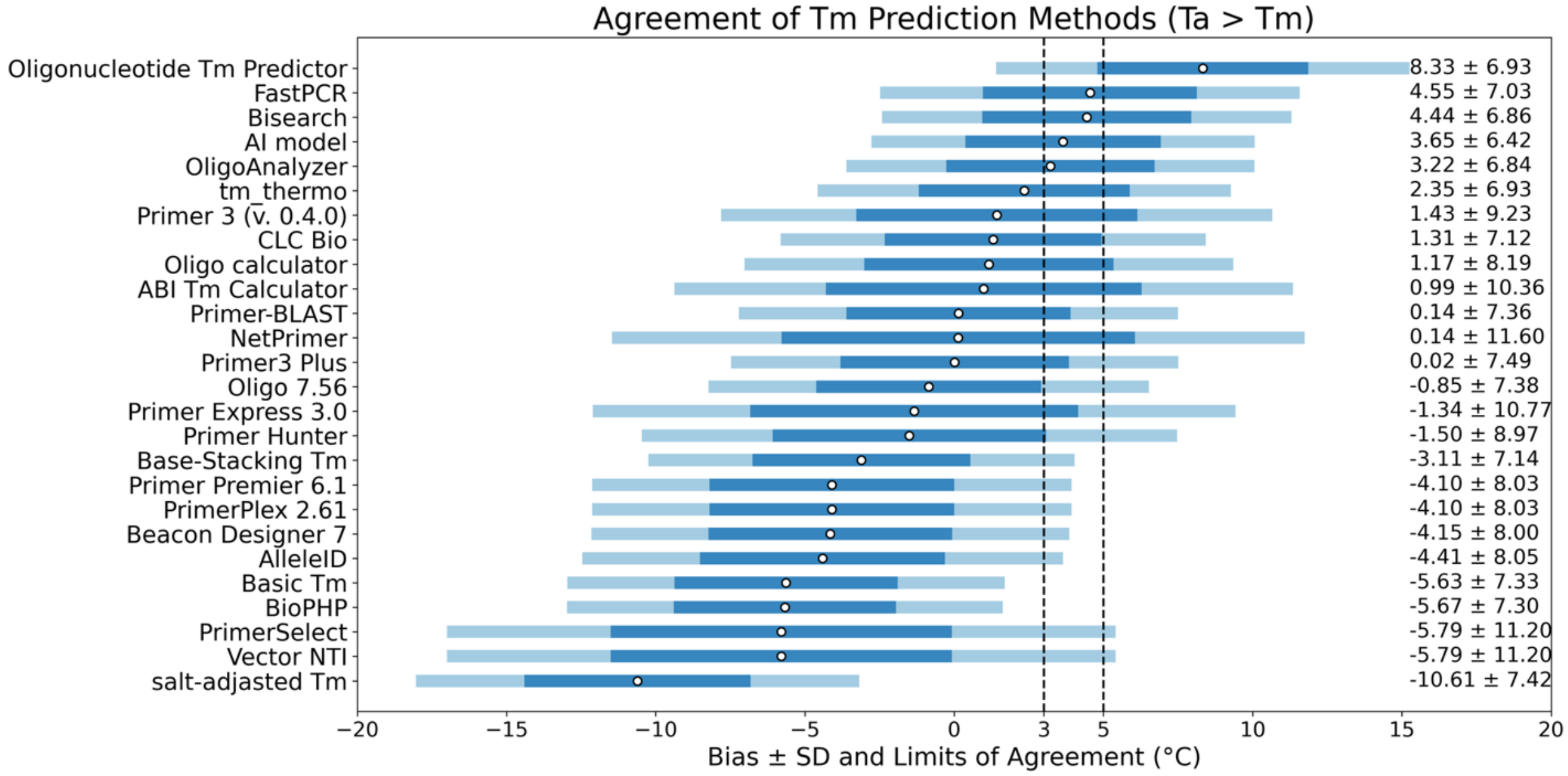

**Figure 7. Agreement with experimental annealing temperatures in the Bakhtiarizadeh benchmark.** Forest plot summarizing Bland-Altman mean bias and 95% limits of agreement (LoA) for the proposed AI model and previously evaluated primer-design tools. Predictions correspond to melting temperature ($T_m$), whereas experimental measurements correspond to annealing temperature (Ta).

Altogether, these external evaluations demonstrate that the proposed framework generalizes beyond the experimental conditions represented during training. The model maintained strong agreement on an independent annealing-temperature benchmark and achieved competitive performance on a DNA thermodynamics benchmark encompassing ionic and concentration regimes substantially outside the training distribution. These results indicate that condition-aware nucleotide language-model representations can retain predictive accuracy under substantial domain shift while remaining competitive with specialized thermodynamic and physics-based approaches.

## Effect of NN-based In-silico Pretraining on Experimental Tm Prediction

To assess the contribution of synthetic thermodynamic pretraining, we compared three variants of the proposed framework trained under different data regimes: (i) Primer3 (P3)-derived synthetic pretraining only, (ii) P3 pretraining followed by experimental fine-tuning, and (iii) experimental training without pretraining (Methods: Synthetic Data Generation). Model predictions were evaluated against experimentally measured $T_m$ values, with Primer3 included as a thermodynamic reference where applicable (Table 4).

Across all experimental datasets, the combination of synthetic pretraining and experimental fine-tuning consistently yielded the strongest performance. For regular oligonucleotides, the fine-tuned model achieved the lowest prediction error (MAE = 0.76 °C) and strongest agreement with experimental measurements ($R^2$ = 0.95), outperforming both the experimental-only model (MAE = 1.16 °C; $R^2$ = 0.89) and the synthetic-only model (MAE = 2.00 °C; $R^2$ = 0.72). Similar trends were observed for both LNA+T and LNA+C datasets, where pretraining followed by fine-tuning reduced MAE by approximately 35-40% relative to training on experimental data alone.

Models trained exclusively on synthetic data performed substantially worse on the experimental datasets despite being derived from established nearest-neighbor thermodynamic principles. This observation suggests that thermodynamically simulated measurements do not fully capture the assay-specific effects and experimental variability present in measured $T_m$ values. Nevertheless, synthetic pretraining provided a useful initialization that, when combined with experimental fine-tuning, consistently reduced error, bias, and limits of agreement across all datasets.

These results indicate that synthetic pretraining and experimental measurements provide complementary information. NN-derived synthetic data encode useful thermodynamic relationships that provide an effective initialization for learning experimentally observed melting behavior, whereas experimental fine-tuning adapts these representations to assay-specific effects and chemically modified nucleotides. Combining both data sources produced the most accurate and robust predictions across unmodified and LNA-modified oligonucleotides.

**Table 4. Effect of synthetic thermodynamic pretraining on $T_m$ prediction performance.**

| Dataset | Method | Training | Bias ↓ | ± | LoA ↓ | MAE (°C) ↓ | RMSE (°C) ↓ | MSE (°C²) ↓ | $R^2$ ↑ | Pearson r ↑ | p-value ↑ |
|---|---|---|---|---|---|---|---|---|---|---|---|
| Regular bases | AI model | **P3 pretrain + Exp train** | **-0.16** | ± | **1.92** | **0.76** | **0.99** | **0.98** | **0.95** | **0.98** | 0.03 |
| | | Exp only (no-pretrain) | 0.24 | ± | 2.82 | 1.16 | 1.46 | 2.12 | 0.89 | 0.95 | 0.007 |
| | | P3 pretrain | 1.75 | ± | 3.05 | 2 | 2.34 | 5.48 | 0.72 | 0.96 | 3.2E-26 |
| | Primer3 (P3) | - | 2.23 | ± | 2.81 | 2.35 | 2.65 | 7.03 | 0.64 | 0.97 | 4.5E-31 |
| LNA+T | AI model | **P3 pretrain + Exp train** | **0.14** | ± | **2.05** | **0.79** | **1.06** | **1.11** | **0.94** | **0.97** | 0.002 |
| | | Exp only (no-pretrain) | 0.47 | ± | 2.86 | 1.25 | 1.53 | 2.34 | 0.88 | 0.95 | 5E-11 |
| | | P3 pretrain | 1.02 | ± | 3.91 | 1.82 | 2.24 | 5.02 | 0.74 | 0.92 | 2.3E-22 |
| LNA+C | AI model | **P3 pretrain + Exp train** | -0.06 | ± | **2** | **0.8** | **1.02** | **1.04** | **0.95** | **0.97** | 0.16 |
| | | Exp only (no-pretrain) | **0.62** | ± | 3.07 | 1.35 | 1.68 | 2.84 | 0.86 | 0.94 | 3.5E-16 |
| | | P3 pretrain | -0.85 | ± | 4.51 | 1.9 | 2.45 | 6 | 0.7 | 0.9 | 7.8E-13 |

The proposed framework was trained using three data regimes: Primer3 (P3) synthetic pretraining only, experimental data only, or P3 pretraining followed by experimental fine-tuning. Metrics include bias, limits of agreement (LoA), MAE, RMSE, $R^2$, Pearson correlation coefficient, and Wilcoxon signed-rank-test p values. Cells are color-scaled by method and metric performance.

# Discussion

Accurate duplex melting temperature ($T_m$) prediction remains a practical challenge for qPCR primer design, diagnostic probes, allele-specific assays, and workflows that incorporate chemically modified oligonucleotides. Classical nearest-neighbor (NN) thermodynamic models have provided the foundation for oligonucleotide design for decades and remain highly effective for canonical DNA duplexes. However, their reliance on fixed parameterizations limits their ability to represent the combined effects of sequence context, assay conditions, and nucleotide modifications. The results presented here suggest that data-driven approaches can complement these thermodynamic foundations by learning context-dependent behavior directly from experimental observations. By combining nucleotide language-model representations with explicit reaction metadata, the proposed framework captures nonlinear sequence- and condition-dependent interactions that are difficult to represent using fixed thermodynamic tables alone, enabling accurate $T_m$ prediction across diverse ionic environments and nucleotide chemistries.

**Practical models from modest experiments.** The strong performance achieved in this study did not arise from experimental data alone, but from the combination of pretrained nucleotide representations, large-scale thermodynamic pretraining, and targeted experimental fine-tuning. Starting from a foundation model trained on genomic sequences and subsequently exposed to millions of synthetic thermodynamic examples, experimental fine-tuning on approximately one thousand oligonucleotide pairs was sufficient to adapt the framework to assay-relevant conditions and LNA-modified chemistries, suggesting that the primary challenge is not acquiring large experimental datasets, but effectively combining prior sequence knowledge, thermodynamic priors, and experimentally measured melting behavior. This potentially lowers the barrier for laboratories seeking to develop $T_m$ predictors tailored to their own buffer compositions, instruments, and modification chemistries.

Despite its use of a foundation-model backbone, the framework remained computationally practical, generating approximately 360 melting-temperature predictions per second during batched inference on a single NVIDIA L40S GPU. This throughput is sufficient for routine oligonucleotide-design and screening workflows.

**Extending thermodynamic modeling through contextual representations.** By combining transformer-based sequence embeddings with explicit reaction metadata ($K^+$, $Mg^{2+}$, oligo concentration), the model learns nonlinear ion–sequence and modification–context interactions that complement the local dinucleotide assumptions of NN models by incorporating broader sequence context and experimentally observed condition-dependent behavior. This enables prediction for modified bases, complex substitution patterns, and buffer regimes that lack hand-curated thermodynamic parameters.

**Improved prediction of modified oligonucleotides.** The most substantial performance gains were observed for LNA-modified oligonucleotides, which represent a particularly challenging class of molecules for thermodynamic modeling. Across both LNA+T and LNA+C chemistries, the proposed framework consistently outperformed the NN-based thermodynamic baseline while maintaining strong agreement with experimental measurements. Improvements were especially evident for terminal and near-terminal modifications, where thermodynamic predictions exhibited larger errors and increased bias, indicating that local sequence context and modification placement contribute to melting behavior in ways that are not fully represented by current NN parameterizations.

**Modification-induced thermal effects ($\Delta T_m$).** Beyond predicting absolute melting temperatures, the model more accurately reproduced $\Delta T_m$ between matched modified and unmodified oligonucleotides. Because oligonucleotide optimization often depends on estimating the thermal consequences of introducing a modification rather than predicting absolute $T_m$ alone, this result highlights the practical utility of the framework for probe and primer design. The improved $\Delta T_m$ further suggest that the model captures context-dependent effects of LNA incorporation that are only partially represented by existing thermodynamic parameterizations. Rather than replacing thermodynamic principles, the framework appears to supplement them with information learned from experimental measurements of modification-dependent behavior.

The pretraining experiments further support this interpretation: synthetic data generated from nearest-neighbor thermodynamic models provided an effective initialization, whereas experimental fine-tuning improved adaptation to real assay conditions and modified nucleotide chemistries.

**Generalization beyond the training distribution.** A key challenge for data-driven approaches is maintaining performance outside the domain represented during training. Despite being trained primarily on assay-like experimental conditions, the model retained strong performance on independent public benchmarks generated under substantially different ionic and concentration regimes. On the Panjkovich-Owczarzy benchmark, which includes potassium concentrations and oligonucleotide concentrations extending far beyond the training distribution, the model achieved performance comparable to established thermodynamic and physics-based approaches. Likewise, on the Bakhtiarizadeh benchmark, the proposed framework exhibited the narrowest limits of agreement among all evaluated prediction tools despite being compared against experimentally determined annealing temperatures rather than melting temperatures.

These results suggest that the framework learns generalizable relationships between sequence, chemistry, and experimental conditions rather than simply memorizing the training distribution, while retaining thermodynamic trends encoded during synthetic pretraining. Notably, this generalization was achieved despite incorporating thermodynamic priors during pretraining, suggesting that the framework builds upon established thermodynamic relationships rather than discarding them. At the same time, the observed differences between external benchmarks and experimental datasets highlight the importance of evaluating predictive models across multiple experimental regimes when assessing real-world utility.

## Limitations and practical tradeoffs

**Data scope and provenance.** Experimental measurements for modified oligos were collected in a single laboratory and cover two LNA substitution types; multi-lab replication and broader chemistry coverage are needed for full generality.

**Scalar target.** The model currently predicts Tm as a single scalar and does not explicitly output thermodynamic parameters (ΔG, ΔH, ΔS) or model folding pathways, secondary structure, or dimerization. For many assay design tasks, a calibrated $T_m$ is sufficient, but mechanistic studies will require richer outputs.

**Extrapolation limits.** Because the experimental $T_m$ distribution is finite (≈52–87 °C), models trained primarily on these data can underperform near distribution edges; in-silico pretraining mitigates but does not eliminate extrapolation risk.

## Practical recommendations and future work

For laboratories prioritizing assay success, we recommend constructing modest, condition-specific training sets and fine-tuning pretrained sequence models with reaction metadata to produce tailored $T_m$ predictors. To broaden applicability, future work should incorporate multi-lab datasets, additional modification chemistries, systematic mismatch and dimer regimes, and hybrid training that mixes assay-like labels with targeted high-salt thermodynamic measurements. Extending the framework to predict full melt-curve representations or joint thermodynamic quantities would increase mechanistic interpretability and utility for tertiary analyses.

## Conclusion

We demonstrate that condition-aware nucleotide language models provide an effective extension of conventional nearest-neighbor thermodynamic approaches for oligonucleotide melting-temperature prediction. By integrating thermodynamic priors, sequence representations with explicit experimental context and leveraging thermodynamic knowledge through synthetic pretraining, the proposed framework achieves sub-degree prediction accuracy across unmodified and LNA-modified oligonucleotides, more accurately captures modification-induced thermal effects, and maintains strong performance on independent external benchmarks. The results further show that reliable and experimentally useful predictors can be developed from relatively modest datasets when combined with modern sequence representations and condition-aware modeling. By prioritizing empirical utility—predicting the melt and the product rather than theoretical values under extreme conditions—we deliver a tool that better matches practitioner needs while remaining extensible to more mechanistic goals.

# Methods

## Oligo Design and Experimental Conditions

Our oligonucleotide dataset was generated using a random-sequence-generator Python script that constrained length to 20–30 nt, reflecting common primer and probe sizes in assay design. Training and test sets were produced with the same procedure and parameter settings.

We additionally created datasets containing LNA-C and LNA-T substitutions. For each modified dataset, the same underlying oligos were used, and C or T bases were replaced in place with the corresponding LNA analog. Up to two C or T positions were substituted per oligo: if an oligo contained fewer than two such bases, only one was modified; if it contained more than two, two positions were selected at random for LNA replacement.

Our goal was to capture the range of conditions most relevant to assay-design practice. To reflect the diversity of Tm-prediction use cases, we evaluated oligos across a broad span of salt conditions ([$K^+$] 0.05–0.5 M; [$Mg^{2+}$] up to 0.011 M) and across oligo concentrations (100-1150 M) representative of typical assay-development workflows.

## Data Preprocessing

Data preprocessing involved deduplication to remove exact duplicates while retaining the highest-quality label. Outlier Tm values inconsistent with reported conditions, such as those below 40 °C, were excluded. Assay condition variables, including magnesium, oligo, and potassium concentrations, were standardized using z-score normalization. Finally, $T_m$ labels were transformed to a logarithmic scale to improve regression stability.

## Experimental $T_m$

Complementary oligonucleotide pairs were purchased from Integrated DNA Technologies (IDT) with standard desalted purity. Upon receipt, oligonucleotides were resuspended in TE buffer (10 mM Tris-HCl, 1 mM EDTA, pH 8.0) in 96-well plates to generate stock solutions. Working dilutions were prepared by aliquoting each oligonucleotide into fresh wells to achieve the desired final concentrations. No PCR or other enzymatic reagents were added; each reaction contained only the complementary oligonucleotides in buffer.

Melt curve analysis was performed using a Thermo Fisher Scientific QuantStudio™ 520 real-time PCR system with Design & Analysis Software v2.8.0. Samples were subjected to the manufacturer's standard melt curve protocol, during which fluorescence was monitored continuously across a programmed temperature ramp. EvaGreen dye was used to detect double-stranded DNA. Melt curves were generated, and $T_m$ were calculated from the first derivative of the fluorescence signal with respect to temperature (–dF/dT) using the instrument's default analysis settings unless otherwise specified. All measurements were performed in technical replicates, and representative Tm values are reported.

## Synthetic Dataset Generation for AI Model Pre-training

To enable large-scale in-silico pretraining, we generated a synthetic $T_m$ dataset using Primer3 to simulate DNA duplex thermodynamics across a broad range of experimentally relevant conditions. A total of 1 million oligonucleotide sequences were generated using a randomized design strategy. All sequences consisted exclusively of standard DNA bases (A, C, G, T). Oligonucleotide lengths were uniformly sampled between 20 and 30 nucleotides (mean ≈ 25 nt), and GC content was controlled to span a wide compositional range (5–95%, centered near 50% GC).

From this sequence library, approximately 3 million in-silico Tm measurements were generated by evaluating each oligonucleotide under multiple combinations of experimental conditions using Primer3. Simulated melting temperatures spanned a broad range (~38–100 °C, mean ≈ 71 °C). Experimental parameters—monovalent salt concentration ($K^+$), divalent magnesium concentration ($Mg^{2+}$), and oligonucleotide concentration—were sampled using a mixture-of-regimes strategy. The majority of samples (~85%) reflected typical PCR-like environments, with $K^+$ concentrations primarily between 100–400 mM, $Mg^{2+}$ between 0–10 mM, and oligonucleotide concentrations between 200–900 nM. A subset (~10%) mimicked conditions commonly found in public thermodynamic datasets, characterized by higher monovalent salt levels (100–1,000 mM), substantially elevated oligonucleotide concentrations (2–100 µM), and no $Mg^{2+}$. The remaining samples (~5%) were drawn across the full allowable parameter ranges ($K^+$: 50–500 mM, $Mg^{2+}$: 0–11 mM, oligonucleotide concentration: 100–1,200 nM) to deliberately probe stress and extrapolation regimes.

## Data Partitioning

For model development, each measurement was represented by an oligonucleotide sequence, seq1, and its corresponding canonical target sequence, seq2, defined as the reverse complement of seq1 after removal of special base additions. Because multiple oligonucleotide variants (regular, LNA-C, or LNA-T) and experimental buffer conditions may correspond to the same underlying target sequence, data were partitioned into training and test sets using a group-aware splitting strategy based on seq2 to prevent sequence leakage, using an 80/20 split at the group level. This ensured that no canonical target sequence appeared in more than one split. The training data were further divided into training and validation subsets using a 85/15 split at the group level, while the test set was held out in its entirety for final evaluation.

We further audited leakage using exact sequence matching, modification-stripped canonical sequences, and reverse-complement-equivalent representations for both seq1 and seq2. No exact, canonical, or reverse-complement overlaps were observed between training and test sets, and no modification-stripped seq1 sequence appeared across multiple splits. Near-neighbor analysis showed that each test sequence was separated from its closest training sequence by at least 6 nucleotides, with a median Levenshtein distance of 10; normalized identity remained low (mean 0.62, maximum 0.74). Synthetic Primer3-derived pretraining sequences were similarly audited against internal and public evaluation sequences, and overlapping records were excluded from the final pretraining corpus. Summary statistics for each split are reported in Table 1, and condition-stratified counts are provided in Supplementary Table S1.

## Sequence Representation and Contextual Feature Integration

Sequences were represented at single-nucleotide resolution with explicit handling of LNA modifications and paired-strand structure. Using the seq1/seq2 definitions above, each input was formatted in a fixed orientation as [CLS] [CONDITION TOKENS] seq1 [SEP] seq2 [EOS], where seq1 is the oligonucleotide/probe strand and seq2 is the canonical target strand. Condition tokens were inserted once immediately after [CLS], before both sequence strands.

The tokenizer was initialized from the pretrained NTv2 vocabulary and extended with [CLS], [SEP], [EOS], and LNA-specific tokens +C and +T. To preserve modified bases as atomic symbols, a deterministic scanner tokenized inputs such that, for example, CTATTCGC+CA+T was represented as [C, T, A, T, T, C, G, C, +C, A, +T]. The embedding matrix was resized to include the extended alphabet. Newly added token embeddings, including +C and +T, were initialized using the model's standard random initializer during embedding resizing and learned during training.

Assay context was provided by token injection rather than hidden-state concatenation. Continuous covariates—oligonucleotide concentration, $[K^+]$, and $[Mg^{2+}]$—were z-score normalized and mapped to numeric condition tokens using a shallow MLP (hidden width = 32, GELU, dropout = 0.1). Categorical condition bins were encoded separately using 16-dimensional embeddings and mapped to the model hidden size. Bin boundaries were: oligonucleotide concentration [−∞, 100, 250, 400, 550, 700, 850, ∞), $[K^+]$ [−∞, 150, 275, 375, ∞), and $[Mg^{2+}]$ [−∞, 2.0, 5.0, 8.0, ∞). This produced two condition tokens per covariate, one numeric and one categorical. Embeddings were projected to the same hidden dimension as the Nucleotide Transformer backbone (768 dimensions) and inserted immediately after the [CLS] token, and the attention mask was extended accordingly, allowing the transformer to attend jointly over condition and sequence tokens.

## AI Model Architecture and Training Procedure

We adapted the pretrained InstaDeepAI/nucleotide-transformer-v2-250m-multi-species[19] backbone for oligonucleotide melting-temperature ($T_m$) prediction. The NTv2 backbone was kept frozen, whereas task-specific adaptation was achieved through trainable low-rank adaptation (LoRA) modules applied to the query, key, value, and dense projections of the transformer layers (rank = 16, $\alpha = 8$, dropout = 0.1). The resulting condition-aware contextual representations were processed by two complementary prediction branches. A global regression head transformed the pooled encoder representation into a sequence-level baseline $T_m$ prediction, capturing broad sequence-context and assay-condition effects. In parallel, local residual heads operated on contextual representations derived from neighboring nucleotide positions. Three local heads were used: a triplet head based on representations of a nucleotide and its immediate 5′ and 3′ neighbors, a 5′ pair head, and a 3′ pair head. These heads modeled position-specific local sequence-context effects, including effects associated with modified bases when present. Local predictions were masked to prevent windows from crossing sequence boundaries, aggregated across valid positions, and incorporated as a residual correction to the global prediction. The final model output was therefore defined as the sum of the global estimate and a scaled local residual contribution.

The global regression head consisted of a linear projection from the pooled 768-dimensional encoder representation to a single scalar output. Each local residual head was implemented as a lightweight residual multilayer perceptron operating on concatenated contextual embeddings from the corresponding local neighborhood. During training, only the LoRA adapters, condition-token embeddings, global regression head, and local residual heads were optimized; all pretrained NTv2 backbone parameters remained fixed.

The model was trained to predict log10($T_m$) by minimizing mean squared error (MSE) on the log-transformed target. For evaluation and reporting, predictions were transformed back to degrees Celsius before computing MAE, RMSE, regression statistics, and Bland-Altman metrics.

Optimization was performed using AdamW with separate parameter groups for the LoRA adaptation parameters, global regression head, and local prediction modules. The base learning rate was $3.11 \times 10^{-5}$, with learning-rate multipliers of 2.0 for the global head and 0.35 for the local heads and condition-token embeddings. Training used a weight decay of 0.01, batch size of 64, gradient clipping at 1.0, and a cosine annealing schedule ($T_{max}$ = 80, $\eta_{min} = 10^{-8}$). The residual local contribution was scaled by a factor of 0.25. Early stopping was applied with a patience of 10 epochs, and checkpoint selection was based on validation-set MSE. Hyperparameters were selected using validation-set performance, while the held-out test set was reserved for final evaluation. Random seeds were fixed to support reproducibility.

To assess whether Primer3-derived synthetic pretraining caused inheritance of Primer3-specific biases, we compared three training regimes: synthetic Primer3 pretraining only, synthetic pretraining followed by experimental fine-tuning, and experimental training without pretraining. Final reported results are based on the synthetic pretraining followed by experimental fine-tuning regime.

## Baselines and Comparative Evaluation

**NN thermodynamic baseline (*Thermo*) -** Model performance was benchmarked against a nearest-neighbor thermodynamic baseline derived from the SantaLucia[6] framework and Owczarzy salt corrections[7]. For each input sequence, duplex enthalpy (ΔH) and entropy (ΔS) were calculated from nearest-neighbor stacking interactions together with terminal AT/GC contributions. The thermodynamic model used a joint parameterization spanning canonical DNA, LNA+C, and LNA+T nearest-neighbor interactions. The parameter vector included DNA and LNA-containing stacking ΔH and ΔS terms, terminal ΔH and ΔS corrections, and monovalent and divalent salt-correction coefficients, yielding 62 free parameters in total. All parameters were estimated simultaneously from the training data using nonlinear least-squares optimization that minimized the squared difference between predicted and experimentally measured melting temperatures. Parameters were initialized from standard literature values where available, and no explicit regularization or parameter constraints were applied. To ensure a fair comparison, the thermodynamic model was fitted using the same training split as the AI framework and evaluated on identical test set.

**Additional baseline models.** To assess the contribution of the proposed architecture, we compared performance against linear regression (LR) and XGBoost models trained using the same sequence- and condition-derived input features. We additionally evaluated Primer3[20] and Visual OMP where applicable. Primer3 was used for comparisons involving unmodified DNA oligonucleotides, whereas Visual OMP was used for evaluating modification-induced thermal effects ($\Delta T_m$) on matched modified and unmodified oligonucleotide pairs.

**External benchmark datasets.** Model generalization was further evaluated using two independent public datasets not used during model development. The Panjkovich-Owczarzy benchmark[12] was used to assess prediction accuracy under ionic and concentration conditions substantially different from those represented in the training data. The Bakhtiarizadeh benchmark[15] was used to compare agreement with experimentally determined annealing temperatures across a collection of primer-design tools. These datasets provide complementary assessments of out-of-distribution performance and real-world applicability.

## Statistical Analysis

All evaluations were conducted on the degree Celsius scale unless otherwise specified. For each oligonucleotide, the difference between experimentally measured $T_m$ and predicted $T_m$ was computed. Model performance was assessed using complementary metrics, including the coefficient of determination ($R^2$), Root Mean Squared Error (RMSE) and Mean Absolute Error (MAE). Uncertainty in MAE and RMSE was quantified using a paired nonparametric bootstrap with $B = 10{,}000$ resamples, resampling oligonucleotide-level pairs $(y_i, \hat{y}_i)$with replacement.

Statistical significance was assessed using two-sided Wilcoxon signed-rank tests on paired prediction–experiment residuals (d = predicted − experimental), evaluating whether the median prediction error differs from zero. Agreement between methods was further assessed using Bland–Altman analysis, reporting mean bias and 95% limits of agreement ($\mu \pm 1.96\sigma$).

Experiments were conducted in an environment with Python 3.10.12, PyTorch 2.8.0 (CUDA 12.8), and Transformers 4.56.1, running on four NVIDIA L40S GPUs (48 GB each). Statistical analyses used SciPy, and performance metrics were computed with scikit-learn.

### Computational Performance

Inference benchmarking was performed on an NVIDIA L40S GPU using batches of 128 oligonucleotide-target pairs. Model inference required approximately 2.9 s for 1,057 examples, corresponding to ~360 predictions per second.

### Data Availability

The external benchmark datasets used in this study are publicly available from https://doi.org/10.1371/journal.pone.0012433.s007 and https://doi.org/10.1371/journal.pone.0012433.s008. The synthetic pretraining dataset generated during this study will be made publicly available upon publication.

## Code availability

Code will be made available upon publication in an appropriate code repository.

# Contributions

D.F., L.L., and M.S. conceived and designed the study. D.F. developed the computational framework, generated the synthetic datasets, performed model training and data analysis., A.A., and N.C. performed the wet-laboratory experiments with guidance from L.L. and R.B., D.F., L.L., and M.S. interpreted the results and wrote the manuscript. All authors reviewed and approved the final manuscript.

## Competing interests

The authors declare no competing interests.

## Additional Information

Correspondence and requests for materials should be addressed to Danielle L. Ferreira. Danielle.lopesferreiraastuto@cepheid.com

# Extended Data

**Supplementary Table S1. Condition-stratified split composition**

| Split | Dataset / chemistry | Oligo concentration | $K^+$ | $Mg^{2+}$ | Tm | Measurements | Unique seq2 |
|---|---|---|---|---|---|---|---|
| Synthetic NN pretrain | regular | 100-99993 | 50-1000 | 0-11 | 38-99 | 2997003 | 1749344 |
| Experimental train | regular | 100-1150 | 50-500 | 0-11 | 52 - 75 | 631 | 631 |
| | lna+c | 100-1000 | 50-500 | 0-11 | 52 - 86 | 1791 | 605 |
| | lna+t | 100-1000 | 50-500 | 0-11 | 52 - 83 | 1813 | 605 |
| Experimental validation | regular | 100-1150 | 50-500 | 0-11 | 58 - 75 | 71 | 71 |
| | lna+c | 100-1000 | 50-500 | 0-11 | 55 - 76 | 199 | 67 |
| | lna+t | 100-1000 | 50-500 | 0-11 | 55 - 77 | 201 | 67 |
| Experimental test | regular | 250-800 | 50-500 | 0-11 | 54 - 83 | 287 | 287 |
| | lna+c | 250-800 | 50-500 | 0-11 | 55 - 83 | 287 | 287 |
| | lna+t | 250-800 | 50-500 | 0-11 | 56 - 83 | 287 | 287 |

**Supplementary Table S2. Position-dependent $T_m$ prediction performance for LNA-modified oligonucleotides.**

| Metric | MAE (°C) ↓ | | | Bias | | |
|---|---|---|---|---|---|---|
| Position | Central | Near-terminal | Terminal | Central | Near-terminal | Terminal |
| AI (+C) | **0.8** | **0.74** | **0.91** | -0.07 | -0.06 | -0.07 |
| Thermo (+C) | 0.93 | 0.94 | 1.29 | -0.22 | 0.17 | 0.98 |
| AI (+T) | **0.7** | **0.92** | **0.83** | 0.28 | 0.04 | -0.05 |
| Thermo (+T) | 0.87 | 1.15 | 1.4 | -0.22 | -0.11 | 1.24 |

Mean absolute error (MAE, °C) and prediction bias (predicted − observed $T_m$, °C) stratified by LNA chemistry and modification position. The AI model maintained lower error and bias across all position classes, with the largest improvements observed for terminal modifications.

## Agreement analysis and Condition-aware behavior (Bland–Altman)

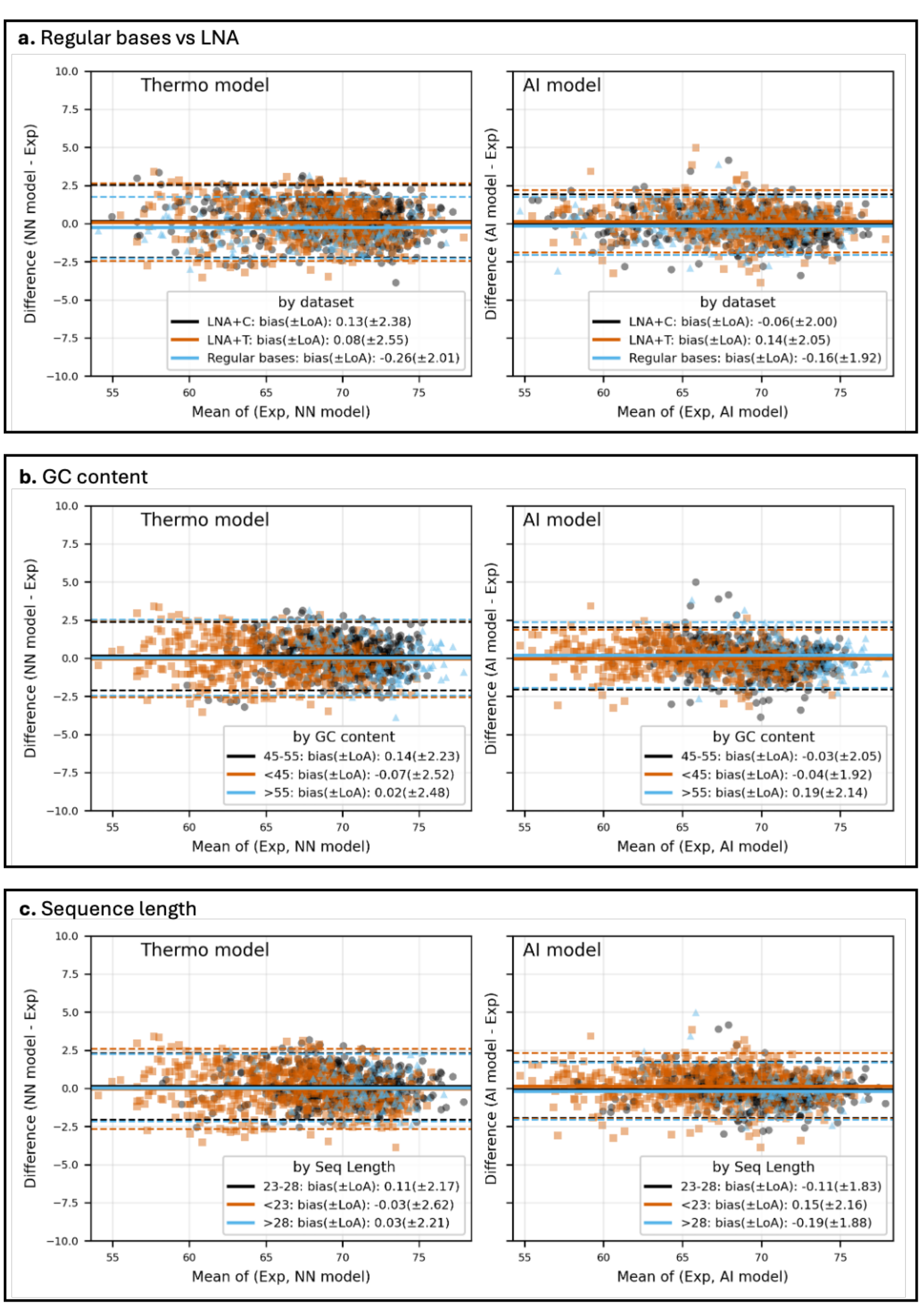

**Figure S1. Agreement analysis across datasets and sequence characteristics.** Bland-Altman plots comparing experimental melting temperatures ($T_m$) with predictions from the thermodynamic baseline (Thermo) and the proposed AI model. Analyses are stratified by (a) dataset, (b) GC-content group, and (c) sequence-length group. Solid lines indicate mean bias and dashed lines indicate the 95% limits of agreement (LoA). Across all strata,

the AI model exhibits narrower limits of agreement and reduced residual variability than the thermodynamic baseline.

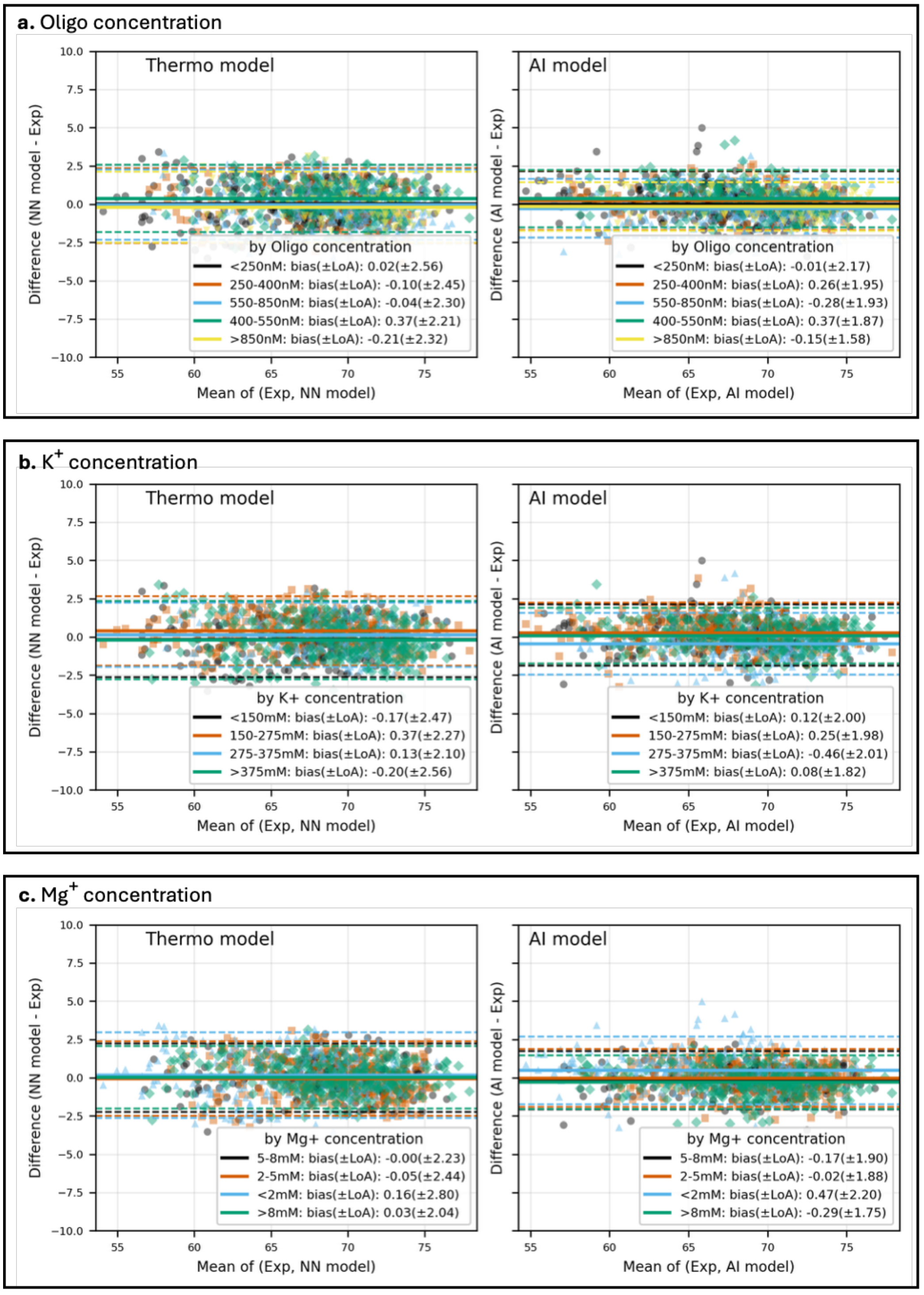

**Figure S2. Agreement analysis across assay conditions.** Bland-Altman plots comparing experimental melting temperatures ($T_m$) with predictions from the thermodynamic baseline (Thermo) and the proposed AI model

across (a) oligonucleotide concentration, (b) potassium ($K^+$) concentration, and (c) magnesium ($Mg^{2+}$) concentration. Solid lines indicate mean bias and dashed lines indicate the 95% limits of agreement (LoA). The AI model maintains stable agreement across a wide range of experimental conditions, with consistently narrower limits of agreement than the thermodynamic baseline.